\documentclass[12pt]{article}

\begin{document}
\title{FINITE DENSITY QCD SUM RULES FOR NUCLEONS}
\author{E. G. Drukarev,
\footnote{present address: PNPI, Gatchina, St. Petersburg 188300,
Russia, e-mail:  drukarev@thd.pnpi.spb.ru}}
\maketitle
\begin{center}
Abstract
\end{center}

It is shown how the QCD sum rules can be applied for the
investigation of the density dependence of the nucleon parameters.
These characteristics can be expressed through the expectation values
of QCD operators in nuclear matter. In certain approximations the
expectation values are related to the observables. First applications
of the approach reproduced some of the basic features of nuclear
physics, providing also a new knowledge. The program of the future work
is presented. The difficulties of the approach are discussed.

\section*{Contents}
\begin{enumerate}
\item {\large Motivation}
\item {\large QCD sum rules in vacuum}

2.1. Dispersion relations\\
2.2. Sum rules

\item {\large QCD sum rules in nuclear matter}

3.1. The problems\\
3.2. Lowest order OPE terms\\
3.3. Gas approximation. The role of $\pi N$ sigma-term in nuclear
physics\\
3.4. Beyond the gas approximation. A possible saturation mechanism\\
3.5. Higher order OPE terms\\
3.6. New knowledge\\
3.7. Self-consistent scenario\\
3.8. A sub-plot: Goldstone pions never condense

\item {\large Summary}
\end{enumerate}

The lecture is not addressed to the experts. The aim of the talk is
rather to attract attention of the researchers who just started to
study the subject. The third section is based on the results obtained
in collaboration with M.~G.~Ryskin, V.~A.~Sadovnikova and E.~M.~Levin.

\section{Motivation}

The theory of nuclear matter leaves some room for the improvement. This
concerns the low densities, i.e. those close to the saturation value as
well as the higher densities.

Since the pioneering paper of Walecka \cite{1} the partially
successful Schr\"odinger phenomenology was succeeded by still more
successful Dirac phenomenology. A nucleon in nuclear matter is treated
as moving in superposition of the scalar and vector fields which are of
several hundreds MeV. In the meson-exchange picture of nucleon
interactions these fields originate from the exchange by $\sigma$ and
$\omega$ mesons. This model is known as quantum hadrodynamics (QHD). It
is quite successful in describing most of the properties of nucleons in
both nuclear matter and finite nuclei \cite{2,3}. However the model is
not fundamentally complete. The weak points of QHD were reviewed by
Negele \cite{4} and by Sliv {\em et al} \cite{5}. Here I mention that
$"\sigma$-meson" is rather an effective way of describing the scalar
interactions. Also the masses $m_{\sigma,\omega}$ of $\sigma$- and
$\omega$-mesons are so large that the scalar and vector interactions
take place at the distances where the nucleons cannot be treated as the
point particles any more. Finally, the coupling constants $g_\sigma$
and $g_\omega$ are the free parameters of QHD. They are chosen usually
to fit the saturation properties.

Thus it would be desirable to develop the approach which has the
attractive features of QHD, avoiding, however, the "meson-exchange"
conception. This would enable us to avoid the controversy of the small
distance description mentioned above.

Another point is connected with the high densities. There are several
interesting phenomena. The chiral phase transition and the
disintegration of the nuclear matter to the quark-gluon plasma are much
discussed nowadays. The possibility of the existence of the other phase
states of nuclear matter containing the admixture of heavier baryons or
of the "pion condensate" have been studied long ago \cite{6,7}. These
effects can be important for the astrophysics. However the QHD
parameters $g_{s,\omega}$ are defined at the saturation point. Hence,
this approach cannot be expanded to the higher densities in a
straightforward way. It is desirable to express the interactions in the
scalar and vector channels through the observables whose density
dependence can be found separately. This would enable to study the high
density nuclear physics.

There are chances that both requirements can be realized. Recall that
the QCD sum rules (SR) method invented by Shifman {\em et al.} \cite{8}
succeeded in expressing the static properties of the hadrons through
the vacuum expectation values of several simplest operators of the
quark and gluon fields (QCD operators of the lowest dimensions). If we
succeed in expanding the SR approach for the case of finite densities,
the in-medium modification of the values of the hadron parameters would
be expressed through the in-medium values of the QCD condensates. Such
approach would not require the conception of heavy meson ($\sigma$ and
$\omega$) exchange. The calculation of the density dependence of QCD
condensates would enable to use the approach in the broad interval of
the density values.

Now I give a brief review of the SR method in vacuum, focusing on the
points which we shall need at the finite values of density. There are
several detailed reviews of the SR approach \cite{9,10}.

\section{QCD sum rules in vacuum}

\subsection{Dispersion relations}

The basic point of SR method is the dispersion relation for the
function $G(q^2)$ which describes the propagation of the system with
the quantum numbers of a hadron. In the simplest form it is
\begin{equation}
G(q^2)\ =\ \frac1\pi \int \frac{\mbox{Im }G(k^2)dk^2}{k^2-q^2}\ .
\end{equation}
In quantum mechanics $G(q^2)$ is just the particle propagator. In the
field theory different degrees of freedom are convenient in the
different regions of the values of $q^2$. In particular, for the system
with the baryon and electric charges  equal to unity $Q=B=1$, the
imaginary part Im$\,G(k^2)=0$ at $k^2<m^2$ with $m$ being the position
of the lowest lying pole, i.e. $m$ is the proton mass. There are the
other singularities at larger values of $k^2$. There are the cuts
corresponding to the systems "proton+pions", etc. On the other hand,
one can consider the system as that of three strongly interacting
quarks. Such description becomes increasingly simple at $q^2\to-\infty$
due to the asymptotic freedom of QCD. This means that at
$q^2\to-\infty$ the function $G(q^2)$ can be presented as the power
series of $q^{-2}$ (and of QCD coupling constant $\alpha_s$). The
coefficients of the expansion are the expectation values of local
operators constructed of quark and gluon fields. These expectation
values are called "condensates". Thus such presentation known as
operator power expansion (OPE) \cite{11}, provides the perturbative
expansion of the short-distance effects, while the nonperturbative
physics is contained in the condensates.

Technically this means that one should start with the general
presentation
\begin{equation}
G(q^2)\ =\ i\int d^4xe^{i(qx)}\langle0|T\{\eta(x)
\bar\eta(0)\}|0\rangle
\end{equation}
with $\eta$ being the local operator with the quantum numbers of the
considered system. The operator $\eta(x)$ is the composition of the
quark fields $\psi(x)$. For each quark field one can write in the
lowest order of $\alpha_s$ (forgetting for a while about colours)
\begin{equation}
\langle
0|T\,q_\alpha(x)\bar q_\beta(0)|0\rangle\ =\ \frac i{2\pi^2}\cdot
\frac{\hat x_{\alpha\beta} +\frac{im_q}2x^2}{x^4}-\frac14\sum_A
\Gamma^X_{\alpha\beta}\,\langle0|:\bar q(0)\Gamma^Xq(x):|0 \rangle
\end{equation}
with $\hat x=x_\mu\gamma^\mu$. This is the direct consequence of Wick
theorem. In Eq. (3) $\alpha$ and $\beta$ are the Lorentz indices, $m_q$
stands for the quark mass. In the second term of the rhs of Eq. (3)
$\Gamma^X$ is the complete set of the basic Dirac $4\times4$ matrices
with the scalar, vector, pseudoscalar, pseudovector and tensor
structures. The first term in the rhs of Eq. (3) is just the free
propagator.

In the theories with the empty vacuum, e.g. in quantum electrodynamics
the second term in rhs of Eq. (3) vanishes due to the normal ordering.
In any field theory all the structures except the scalar one vanish due
to the Lorentz invariance. In QCD the scalar term survives due to the
spontaneous breaking of the chiral  invariance. Thus
\begin{equation}
\langle0|Tq^a_\alpha\,(x)\bar q^b_\beta(0)|0\rangle\ =\ \frac
i{2\pi^2}\, \frac{\hat
x_{\alpha\beta}+(im_q)/2\,x^2}{x^4}\,\delta^{ab}-\frac1{12}\,
\langle0|\bar q^a(0)q^a(x)|0\rangle
\end{equation}
with $"a"$ and $"b"$ standing for the colour indices. Note that Eq.(4)
has a simple physical meaning. The quark can propagate between the
space-time points $"0"$ and $"x"$ as a free particle or by the exchange
with the vacuum sea of the quark-antiquark pairs. Note, however, that
Eqs. (3) and (4) are not presented in a gauge-invariant way (the
operator $q(x)$ depends on the gauge of the gluon fields). The rigorous
and gauge-invariant form of Eq. (4) is
\begin{equation}
\langle 0|Tq^a_\alpha(x)\bar q^b_\beta(0)|0\rangle\ =\ \frac i{2\pi^2}
\, \frac{\hat x_{\alpha\beta}+i(m_q)/2\,x^2}{x^4}\,
\delta_{ab}-\frac1{12}\,\langle0|\bar q^a(0)q(0)|0\rangle\, +0(x^2)
\end{equation}
with the expansion in powers of $x^2$ corresponding to the expansion of
$G(q^2)$ in powers of $q^{-2}$.

To demonstrate the power of the dispersion relations I present,
following \cite{10}, the derivation of the well-known
Gell-Mann, Oakes--Renner relation (GMOR) \cite{12}
\begin{equation}
\langle 0|\bar uu+\bar dd|0\rangle\ =\ -\ \frac{2f^2_\pi
m^2_\pi}{m_u+m_d}
\end{equation}
with $f_\pi$ and $m_\pi$ being the decay
constant and the mass of the pion. Recall that Eq. (6) is true in the
chiral limit $m^2_\pi\to0$, $m_q\to0$.

The quantum numbers of pion can be carried by the axial current
$A_\mu(x) =\bar u(x)\gamma_\mu\gamma_5d(x)$ as well as by the
pseudoscalar current $P(x)=i\bar u(x)\gamma_5d(x)$. Consider the
dispersion relation for the function
\begin{equation}
G(q^2)\ =\ i\ \frac{q^\mu\int d^4x\,e^{i(qx)} \langle0|TA_\mu(x)\bar
P(0)|0\rangle}{q^2}
\end{equation}
using Eq. (5) for the quark propagators. The corresponding integral
diverges at small $x$. Introducing a cutoff $x^2\ge L^2$ one finds in
the limit $\alpha_s=0$
\begin{equation}
G(q^2)\ =\ i(m_u+m_d)\ \frac3{8\pi^2}\ln\frac{L^2}{-q^2}+i\
\frac{\langle0|\bar uu+\bar dd|0\rangle}{q^2}\ .
\end{equation}
To obtain the rhs of Eq. (1) recall about the partial conservation by
axial current (PCAC) expressed by the equation $D^\mu A_\mu(x)=\sqrt2
f_\pi m^2_\pi\varphi(x)$ with $\varphi$ standing for the pion field.
Present
\begin{equation}
\mbox{Im }G(k^2)\ =\ \langle0|A_\mu k^\mu|\pi(k)\rangle\langle\pi(k)
|\bar P|0\rangle\,\delta(k^2-m^2_\pi)+R(k^2)
\end{equation}
with the term $R(k^2)$ describing the higher lying states. Using PCAC
one finds the term $R(k^2)$ to contain one more factor $m^2_\pi$
compared to the first term in rhs of Eq. (9). Hence, in the chiral
limit we can neglect $R(k^2)$ as well as the first in rhs of Eq. (8).
Thus the dispersion relation is
\begin{equation}
i\ \frac{\langle0|\bar uu+\bar dd|0\rangle}{q^2}\ =\ \frac{\langle0
|A_\mu k^\mu|\pi\rangle\langle\pi|\bar P|0\rangle}{m^2_\pi-q^2}\ ,
\end{equation}
with $k^2=m^2_\pi$.
Calculating the matrix elements in the rhs by using PCAC and assuming
$-q^2\gg m^2_\pi$ we come to Eq. (6).

This derivation of GMOR relation is complementary to the standard one
presented in the QCD textbooks (see, e.g. \cite{13}) which is based on
physics of small momenta $q$. Usually GMOR is treated as the way to
determine the value of $\langle0|\bar uu+\bar dd|0\rangle$. However in
the framework of the developed approach it can be viewed as the
relation which expresses the combination of the pion parameters
$m^2_\pi f^2_\pi$ through the expectation value
$\langle0|\bar uu+\bar dd|0\rangle$.

\subsection{Sum rules}

Unfortunately, the example considered above is the only case when the
dispersion relation takes such a simple form. Usually there is no
reason to neglect the higher lying physical states with respect to the
lowest one. If the second lowest singularity is the cut starting at the
point $W^2_{ph}$ we can present
\begin{equation}
\mbox{Im }G(k^2)\ =\ \lambda^2\delta(k^2-m^2)+f(k^2)\theta
(k^2-W^2_{ph})
\end{equation}
with $\lambda^2$ being the residue while $f(k^2)$ is the spectral
function. Following previous discussion the lhs of Eq. (1) can be
expanded in powers of $q^{-2}$ and thus Eq. (1) takes the form
\begin{equation}
G_{OPE}(q^2)\ =\
\frac{\lambda^2}{m^2-q^2}+\frac1\pi\int\limits_{W^2_{ph}}
\frac{f(k^2)dk^2}{k^2-q^2}
\end{equation}
with the unknown parameters $m,\lambda^2$, and the unknown spectral
function $f(k^2)$. The aim of the SR is to obtain the parameters of the
lowest lying state. Hence, the second term of rhs of Eq. (12) is
treated approximately. The approximation is prompted by the asymptotic
behaviour
\begin{equation}
f(k^2)\ =\ \frac1{2i}\ \Delta\,G_{OPE}(k^2)
\end{equation}
at $k^2\gg|q^2|$. The integral over $k^2\gg|q^2|$ provides the terms
$\sim\ln(L^2/-q^2)$ exceeding the contribution of the pole by this
factor. The standard ansatz consists in extrapolation of Eq. (12) to
the lower values of $k^2$, replacing also the physical threshold
$W^2_{ph}$ by the unknown effective threshold $W^2$, i.e.
\begin{equation}
\frac1\pi\int\limits^\infty_{W^2_{ph}} \frac{f(k^2)}{k^2-q^2}\,dk^2\ =\
\frac1{2\pi^2i}\int\limits^\infty_{W^2}
\frac{\Delta\,G_{OPE}(k^2)}{k^2-q^2}\ dk^2\ ,
\end{equation}
and thus the dispersion relation (1) takes the form
\begin{equation}
G_{OPE}(q^2)\ =\ \frac{\tilde\lambda^2}{m^2-q^2} +
\frac1{2\pi^2i}\int\limits^\infty_{W^2}
\frac{\Delta\,G_{OPE}(k^2)}{k^2-q^2}\ dk^2\ .
\end{equation}
Such approximation of the spectrum is known as the "pole + continuum"
model.

The lhs of Eq. (15) contains the QCD condensates.  The rhs contains
three unknown parameters $m,\lambda^2$ and $W^2$. However both lhs and
rhs depend on $q^2$. The OPE becomes increasingly true at large values
of $-q^2$. The "pole + continuum" model has sense only if the
contribution of the continuum, treated approximately does not exceed
the contribution of the pole, treated exactly. Thus the model becomes
increasingly true at small values of $|q^2|$. The problem is to find
the region of $|q^2|$ where both OPE and "pole + continuum" model are
valid.

Such region  is unlikely to exist in any channel of the dispersion
relations presented by Eq. (15) with the necessary subtractions. To
improve the overlap between the QCD and phenomenological descriptions
Borel transform defined as
\begin{eqnarray}
Bf(Q^2) &=& \lim\limits_{Q^2,n\to\infty} \frac{(Q^2)^{n+1}}{n\ !}\left(
\frac{-d}{dQ^2}\right)^n f(Q^2)\ \equiv\ \bar f(M^2) \\
&& \hspace*{3.cm} Q^2\ =\ -q^2\ ; \quad M^2\ =\ Q^2/n \nonumber
\end{eqnarray}
was used in \cite{8}. There are several useful features of the Borel
transform. It removes the divergent terms in the lhs of Eq. (15) which
are caused by the free quark loops --- see, e.g., Eq. (8). This
happens, since the Borel  transform eliminates all the polynomials in
$q^2$. Thus we do not need to make subtractions. Also it emphasizes
the contribution of the lowest lying states in rhs of Eq. (15) due to
the relation
\begin{equation}
B\ \frac1{Q^2+m^2}\ =\ e^{-m^2/M^2}\ .
\end{equation}

The Borel transformed dispersion relations
\begin{equation}
\tilde G_{OPE}(M^2)\ =\ \lambda^2e^{-m^2/M^2}+\frac1{2\pi i}
\int\limits^\infty_{W^2} dk^2 e^{-k^2/M^2}\cdot\Delta\,G_{OPE}(k^2)
\end{equation}
have been analyzed first for the vector mesons \cite{8} and for the
nucleons \cite{14}. In both cases the regions of the values of the
Borel mass $M$ were found, where the matching of rhs and lhs of Eq.
(18) was achieved. In particular, Ioffe \cite{14} calculated the value
of the meson mass as the function of QCD condensates. Several terms of
OPE appeared to be needed to provide the matching of rhs and lhs of Eq.
(18). The result of \cite{14} can be treated as
\begin{equation}
m\ =\ C\,\langle0|\bar uu|0\rangle
\end{equation}
with $C<0$ being the function of the gluon condensate and of the
four-quark condensate $\langle0|\bar qq\bar qq|0\rangle$. The latter
can be viewed as the expansion of the two-quark propagator, similar to
Eq. (3). One can see that Eq. (19) has a simple physical meaning. The
nucleon mass is caused by the quark exchange with the vacuum sea of
$\bar qq$ pairs. The mechanism resembles that of the Nambu and
Jona--Lasinio model.

The QCD SR method was applied successfully to calculation of the static
properties of mesons \cite{8} and nucleons \cite{14,15}. It provided
new knowledge as well. For example, the value of gluon condensate
$g_0=\langle0|\frac{\alpha_s}\pi G_{\mu\nu}G_{\mu\nu}|0\rangle$ was
extracted by Vainshtein {\em et al.} \cite{16} from the analysis of
leptonic decays of $\rho$ and $\varphi$ mesons and from QCD analysis of
charmonium spectrum. This condensate is a very important
characteristics of QCD vacuum, since it is directly related to the
vacuum energy density. The investigations based on the vacuum QCD sum
rules are going on until now. The latest HEP preprint \cite{17} which
the value of $g_0$ has been calculated more accurately was published
several months ago.

\section{QCD sum rules in nuclear matter}

\subsection{The problems}

Now we discuss the possibility of the extension of QCD SR approach to
the investigation of the characteristics of the nucleons in nuclear
matter. If we succeed, the modification of the characteristics of the
nucleon will be expressed through the expectation values of QCD
operators in medium. Although some of the qualitative results may find
the applications in the investigations of the finite nuclei, only the
infinite nuclear matter will be considered below. In other words, the
density of the distribution of the nucleons is the same in all the
space points.

Since the Lorentz invariance is lost, the correlation function in
medium
\begin{equation}
G^m(q)\ =\ i\int d^4xe^{i(qx)}\ \langle M|T\left\{\eta(x)\bar\eta(0)
\right\}|M\rangle
\end{equation}
depends on two variables, but not on $q^2$ only. The spectrum of the
function $G^m(q)$ is much more complicated, than that of the vacuum
correlator $G(q^2)$ defined by Eq. (2). The singularities can be
connected with the nucleon (proton) placed into the matter, as well as
with the matter itself. One of the problems is to find the proper
variables,  which would enable us to focus on the properties of our
probe proton.

In the papers \cite{18,19} it was suggested to use $q^2$ as one of the
variables. The shift of the position of the nucleon pole $m_m-m$ would
be the unknown parameter to be determined from the SR equations. On the
other hand
\begin{equation}
m_m-m\ =\ U\left(1+0\left(\frac Um\right)\right)
\end{equation}
with $U$ being the single-particle energy of the nucleon. This is the
very characteristics which enters the equation of state. To separate
this singularity from the other ones a proper choice of the second
variable is needed. Considering the nuclear matter as the system of $A$
nucleons with momenta $p_i$, introduce
\begin{equation}
p\ =\ \frac{\Sigma p_i}A
\end{equation}
with ${\bf p}=0$ in the rest frame of the matter. Under the choice of
$s=(p+q)^2=$const we avoid the singularities connected with the
excitation of two nucleons \cite{18}--\cite{20}. The constant value of
$s$ should be fixed by the condition that the probe proton is put on
the Fermi surface of the matter. In the simplified case when the Fermi
motion of the nucleons of the matter is neglected, we can just assume
$s=4m^2$.

Thus we shall write the dispersion relations for the function
$G^m(q^2,s)-G(q^2)$ with the functions $G$ and $G^m$ defined by Eqs.
(2) and (20). It was shown in \cite{21} that the nucleon pole is still
the lowest lying singularity of the function $G^m(q^2,s)$ until we do
not include the three-nucleon interactions. All the other singularities
are lying at larger values of $q^2$ being quenched by the Borel
transform due to Eq. (17). Thus we use the "pole + continuum" model for
the spectrum of the function $G^m(q^2,s)$.

The OPE coefficients of $G^m(q^2,s)$ are the in-medium expectation
values of QCD operators. Thus to use the SR equations one should find
the density dependence of these condensates. It is not clear
{\em"a~priori"} if the OPE series  converges indeed.

\subsection{Lowest order OPE terms}

In the lowest order of OPE the expectation values of the lowest
dimension are involved only. In particular, there are the scalar
expectation values
\begin{equation}
\kappa^i(\rho)\ =\ \langle M|\bar q_iq_i|M\rangle
\end{equation}
with $q_i$ standing for $"u"$ or $"d"$ quark. There are also the vector
expectation values
\begin{equation}
v_\mu^i(\rho)\ =\ \langle M|\bar q_i\gamma_\mu q_i|M\rangle
\end{equation}
taking the form $v_\mu^i(\rho)=v^i(\rho)\delta_{\mu0}$ in the rest
frame of the matter. The condensates $\kappa$ have nonzero values in
vacuum, while the vacuum values of the vector condensates vanish, i.e.
$v_i(0)=0$. Due to the conservation of the vector current we find
immediately
\begin{equation}
v^i(\rho)\ =\ \frac{n_i^{(p)}+n_i^{(n)}}2\,\rho
\end{equation}
with $n^{p(n)}_i$ standing for the number of the valence quarks of the
flavour $"i"$ in the proton (neutron).

The correlator $G^m(q^2)$ contains three structures being proportional
to $\gamma_\mu p^\mu$, $\gamma_\mu q^\mu$, and $I$ with $I$ standing
for the unit $4\times4$ matrix. Thus we obtain three QCD sum rules.
There are four independent unknowns to be determined from SR equations.
These are the three parameters of the nucleon which are the vector and
scalar self-energies $\Sigma_s$ and $\Sigma_s$ and also the shift of
the value of the residue $\lambda^2_m-\lambda^2$. The shift of the
position of the pole can be expressed through the self-energies, i.e.
$m_m-m=\Sigma_v+\Sigma_s$. One more unknown parameter is the shift of
the position of the threshold $W^2_m-W^2$.

The explicit form of the SR equations is presented, e.g. in \cite{21}.
The matching of the rhs and lhs can be achieved in the same interval of
the values of $M^2$ as in the case of vacuum. The shift of the position
of the nucleon pole was found to be a superposition of the vector and
scalar condensates \cite{20}
\begin{eqnarray}
&& m_m\ =\ C_1\kappa(\rho)+C_2v(\rho) \nonumber\\
&& m_m-m\ =\ U(\rho)
\end{eqnarray}
with the last equality coming form Eq. (21),
$\kappa=\kappa^u+\kappa^d$. This provides the simple picture of
formation of the value of $m_m$. Our probe proton exchanges quarks with
the sea of $\bar qq$ pairs which differs from that in vacuum. This
forms the Dirac effective mass $m^*=m+\Sigma_s$. This mechanism is
described by the first term in rhs of Eq. (26). The exchange with the
valence quarks adds the second term.

In the next to leading order of OPE the gluon condensate
$g(\rho)=\langle M|\frac{\alpha_s}\pi G_{\mu\nu}|M\rangle$ should be
taken into account. We shall see that numerically it is not very
important.

\subsection{Gas approximation. The role of $\pi N$ sigma-term in
nuclear physics}

We saw the lowest OPE to contain the condensates $v(\rho)$,
$\kappa(\rho)$ and $g(\rho)$. While the vector condensate is exactly
linear in $\rho$, the condensates $\kappa(\rho)$ and $g(\rho)$ are more
complicated functions of density. We start with the gas approximation
in which the matter is treated as ideal Fermi gas of the nucleons. Thus
our probe proton interacts with the system of non-interacting nucleons.
In this approximation \cite{18}
\begin{eqnarray}
\kappa(\rho) &=& \kappa(0)+\rho\langle N|\bar qq|N\rangle\ ,\\
g(\rho) &=& g(0)+\rho\langle N|\frac{\alpha_s}\pi\,
G_{\mu\nu}G_{\mu\nu}|N\rangle\ .
\end{eqnarray}

The matrix elements in rhs of Eqs. (27) and (28) can be related to the
observables. In particular,
\begin{equation}
\langle N|\bar qq|N\rangle\ =\ \frac{2\sigma}{m_u+m_d}
\end{equation}
with $\sigma$ being the pion-nucleon sigma term which is connected to
the pion-nucleon elastic scattering amplitude $T$ by the relation
$T=-\sigma/f^2_\pi$ \cite{22}. However in the latter relation $T$
denotes the amplitude in certain unphysical point. The experiments
provide the data on the physical amplitude $T_{ph}=-\Sigma/f^2_\pi$
with $\Sigma=(60\pm7)\,$MeV. The method of extrapolation of observable
amplitude to the unphysical point was developed by Gasser {\em et~al.}
\cite{23}. They found
\begin{equation}
\sigma\ =\ (45\pm7)\mbox{ MeV }.
\end{equation}
Note that from the point of chiral expansion, the difference
$\Sigma-\sigma$ is of higher order, i.e. $(\Sigma-\sigma)/\sigma\sim
m_\pi$.

Note also the physical meaning of the expectation value of the operator
$\bar qq$ averaged over a hadron state. Anselmino and Forte
\cite{24,25} showed that under reasonable model assumptions it can be
treated as the total number of quarks and antiquarks.

As to gluon condensate, the expectation value is \cite{26}
\begin{equation}
\langle N|\frac{\alpha_s}\pi\,G_{\mu\nu}G_{\mu\nu}|N\rangle\ =\
-\frac89\bigg(m-\Sigma m_j\cdot\langle N|q_jq_j|N\rangle\bigg)
\end{equation}
with $j$ standing for $u,d$ and $s$ quarks. This equation comes from
the averaging of the QCD Hamiltonian over the nucleon state with the
account of the additional relations found in \cite{26}.
In the chiral limit only
the strange quarks contribute. In the chiral SU(3) limit the second
term in brackets in rhs of Eq. (31) turns to zero.

Solving the SR equations in the leading order of OPE in the gas
approximation we find the potential energy to be \cite{20}
\begin{equation}
U(\rho)\ =\ bigg[66v(\rho)-32(\kappa(\rho)-\kappa(0))\bigg]
\mbox{ GeV}^{-2}
\end{equation}
with the difference $\kappa(\rho)-\kappa(0)$ being described by the
second term of the rhs of Eq. (27). Thus the potential energy is
presented as the superposition of a positive term proportional to the
vector condensate and a negative term proportional to the scalar
condensate. At the saturation point $\rho=\rho_0=0.17\,\rm
fm^{-3}=1.3\cdot10^{-3}GeV^3$ we find the two terms in the rhs of Eq.
(32) to be 200 MeV and -330 MeV. The gluon condensate adds about 10 MeV
to the vector term in the next to leading order of OPE Similar results
were obtained in \cite{27} in another SR approach based on the
dispersion relations in $q_0$.

These are the common points between the SR approach and the Walecka
model \cite{20}. Note, however, that we did not need the fitting
parameters like $g_\omega$ and $g_\sigma$ of QHD. The exchanges by the
strongly correlated quarks (i.e. by the mesons) are expressed through
the exchanges by the uncorrelated quarks. The interactions in the
vector channel are calculated explicitly. The interactions in the
scalar channel are expressed through the observable $\pi N$ sigma-term.
Hence, in SR approach the $\sigma$-term  determines the linear part of
the scalar interactions.

\subsection{Beyond the gas approximation. A possible saturation
mechanism}

Now we shall try to go beyond the gas approximation, remaining,
however, in the lowest orders of OPE. Account of the interactions
between the nucleons of the matter does not change Eq. (25) for the
vector condensate. However the scalar condensate obtains additional
contributions $S(\rho)$ caused by averaging of the operator $\bar qq$
over the meson cloud. Thus we have
\begin{equation}
\kappa(\rho)\ =\ \kappa(0)+\rho\langle N|\bar qq|N\rangle +S(\rho)
\end{equation}
with a nonlinear behavior of $S(\rho)$ at small values of $\rho$.

Assume that the meson cloud consists of all kinds of the mesons
$(\pi,\omega$, {\em etc.}). It was shown in \cite{18} that in the chiral
limit $m^2_\pi\to0$ (neglecting also the finite size of the nucleons)
one can obtain the function  $S(\rho)$ as a power series in Fermi
momenta $p_F\sim\rho^{1/3}$. The lowest order term $\sim\rho p_F$ comes
from the one-pion Fock term (known also as the Pauli blocking term).
The two-pion exchanges with the nucleons in the intermediate states
contribute as $\rho p^2_F$ \cite{28}. The heavier mesons contribute as
$\rho p^3_F\sim\rho^2$.

Even beyond the chiral limit we expect the pion cloud to provide the
leading contribution to the nonlinear term $S(\rho)$. This is because
the contributions of various mesons $X$ contains the meson expectation
values $\langle X|\bar qq|X\rangle=n_X$ with $n_X$ being the total
number of quarks and antiquarks. We can expect $n_X\approx2$. However
the pion expectation value calculated by the current algebra technique
\cite{29} provides $\langle\pi|\bar
qq|\pi\rangle=\frac{m_\pi}{m_u+m_d}\approx12$. Thus the pion cloud
contribution is enhanced.

Note that the simplest account of the nonlinear terms in the scalar
condensate signals the possible saturation mechanism.  Assuming the
chiral limit $m^2_\pi=0$ and including the Pauli blocking term only we
can present
$\kappa(\rho)=\kappa(0)+\frac{2\Sigma}{m_u+m_d}\rho-3.2
\frac{p_F}{p_{F0}}\rho$
\cite{21} with $p_{F0}=268\,$MeV/c being the Fermi momentum corresponding
to the saturation value of density $\rho_0$. This provides the potential
energy
\begin{equation}
U(\rho)\ =\ \left[\left(198-42\cdot\frac{2\Sigma}{m_u+m_d}\right)
\frac\rho{\rho_0}+133\left(\frac\rho{\rho_0}\right)^{4/3}\right]\mbox{MeV
},
\end{equation}
which contains the $\Sigma$-term and the pion-nucleon coupling constant
$g_{\pi NN}$ as the only parameters. After adding the kinetic energy
the energy functional obtains the minimum at $\rho=\rho_0$ if we put
$\Sigma=62.8\,$MeV ($\sigma=47.8\,$MeV) which is consistent with the
experimental data. The binding energy appears to be
$\varepsilon=-9\,$MeV. The incompressibility coefficient which defines
the shape of the saturation curve also has a reasonable value
$K\approx180\,$MeV.

Of course, the results for the saturation should not be taken too
seriously. They are very sensitive to the exact value of the $\Sigma$
term. This is caused by the simplified model of the nonlinear effects.
The rigorous treatment of the pion cloud requires the account of the
multinucleon effects in the propagation of the pions \cite{30,31}.
Inclusion of these effects \cite{32,33} still provides $S(\rho)<0$ at
the densities close to the saturation value. Thus the nonlinear
behaviour of the scalar condensate may be responsible for the
saturation properties of the matter. However the results of this
subsection can be considered only as the sign that further development
of the approach may appear to be fruitful.

Thus the nonlinear behaviour of the scalar condensate is a possible
source of the saturation. Here we find a certain analog of the QHD
saturation mechanism. Recall that in Walecka model the saturation caused
by a complicated dependence of the "scalar density" of the nucleons
$\rho_S=\int\frac{d^3p}{(2\pi)^3}\frac{m^*}{\varepsilon(p)}\theta(p_F
-p)$ on the density $\rho=\int\frac{d^3p}{(2\pi)^3}\theta(p_F-p)$.

Anyway, further development of the approach requires investigation of
the higher order OPE terms as well as the analysis of the condensates
beyond the gas approximation.

\subsection{Higher order OPE terms}

As we have seen, the vector and scalar condensates only contribute in
the leading order of OPE. The gluon condensate contributes to the terms
of the relative order $q^{-2}$ of the OPE of $G^m(q^2,s)$. We saw it to
be numerically small, as well as some other contributions of this order
which we did not discuss (see \cite{33}). However there are some
problems with the terms of the order $q^{-4}$ which contain the
four-quark condensates $h^{XY}(\rho)=\langle M|\bar q\Gamma^Xq\bar
q\Gamma^Yq|M\rangle$.

The expectation values of the four-quark operators are not well
established even in vacuum. The usual assumption is the factorization
approximation \cite{8} with the intermediate vacuum states dominating
in all the channels. As to the in-medium values of the scalar
condensate, one can separate the configuration with one of the products
$\bar qq$ acting on vacuum states while the other one acts on the
nucleon states \cite{20}. Thus in the gas approximation
\begin{equation}
h^{SS}(\rho)-h^{SS}(0)\ =\ 2\langle0|\bar qq|0\rangle\langle N|\bar
qq|N\rangle\rho+\langle N|(\bar qq\,\bar qq)_{int}|N\rangle\rho\ .
\end{equation}
In the second term of the rhs all the quark operators act inside the
nucleon. Equations (6) and (29) enable us to find the magnitude of the
first term of the rhs. If the first term estimates the values of all
the four-quark condensates indeed, the term of the order $q^{-4}$ of
OPE is numerically larger than the leading term at the
values of the Borel mass where the SR equations are solved.  This would
cause doubts in the convergence of OPE.  Fortunately the situation is
not as bad as that. Celenza {\em et~al.} \cite{34} demonstrated that
there is a strong cancellation between the two terms in rhs of Eq.
(34).  However the lack of information on the four-quark condensate
have been an obstacle for the further development of the approach
during many years.  The recent calculations \cite{35} are expected to
improve the situation.

\subsection{New knowledge}

Independently of the magnitude of the contribution of the four-quark
condensates, the SR predict certain new features of the nuclear forces.

\subsubsection{Anomalous structure of the nucleon-meson vertices}

As we have stated earlier, the QCD sum rules can be viewed as a
connection between exchange of uncorrelated $\bar qq$ pairs between our
probe nucleon and the matter and the exchange by strongly correlated
pairs with the same quantum numbers (the mesons). In the conventional
QHD picture this means that in the Dirac equation for the nucleon in
the nuclear matter $(\hat q-\hat V)\psi=(m+\Phi)\psi$ the vector
interaction $V$ corresponds to exchange by the vector mesons with the
matter while the scalar interaction $\Phi$ is caused by the scalar
mesons exchange. In the mean field approximation the vector interaction
$V$ is proportional to density $\rho$, while the scalar interaction is
proportional to the "scalar density" $\rho_s$ which is a more
complicated function of the density $\rho$. Thus $V=V(\rho)$, while
$\Phi=\Phi(\rho_s)$. We have seen that QCD sum rules provide similar
picture in the lowest orders of OPE: vector and scalar parts $G_{v,s}$
of the correlator $G^m$ depend on vector and scalar condensates
correspondingly: $G^m_v=G^m_v(v(\rho))$; $G^m_s=G^m_s(\kappa(\rho))$.
However we find a somewhat more complicated dependence in the higher
order OPE terms, say, $G^m_s=G^m_s(\kappa(\rho),v(\rho))$, depending on
both scalar and vector condensates. This is due to the four-vector
condensates. In particular, the scalar-vector condensate $\langle
M|\bar uu\bar d\gamma_0d|M\rangle$ contains the contribution
$\langle0|\bar uu|0\rangle\langle M|\bar d\gamma_0d|M\rangle$. This
term is proportional to the vector condensate, contributing, however,
to the SR for the scalar structure. In QHD picture this corresponds to
the explicit dependence of the scalar interaction $\phi$ on the density
$\rho$. Such dependence is not included in QHD, at least on the mean
field level. These contributions correspond to the anomalous structure
of vertex of the interaction between the nucleon and the scalar meson
in the meson exchange picture. Similar situation takes place for the
condensate $\langle0|\bar uu|0\rangle\langle M|\bar uu|M\rangle$
contributing to the SR for the vector structure.

\subsubsection{Charge-symmetry breaking forces}

There is an old problem of the difference between the strong
interactions of the proton and neutron with the systems containing
equal numbers of protons and neutrons \cite{36}. In framework of SR
method the difference can be caused by the explicit dependence of the
function $G^m$ on the current quark masses and on the isospin breaking
expectation value $\gamma=\frac{\langle M|\bar dd-\bar uu|M\rangle}{
\langle M|\bar uu|M\rangle}$. The problem was attacked by the SR
approach in a number of works \cite{37}. The proton-neutron binding
energy difference was expressed through the quark mass difference
$m_d-m_u$ and the condensate $\gamma$ under various additional
assumptions. The SR analysis provided some qualitative results which
may be useful in the building of the charge-symmetry breaking nuclear
forces (CSB). One of them is the importance of CSB in the scalar
channel. Earlier there was a common belief that the CSB in the vector
channel are responsible for the effect. Another point is that the mixed
structures described in Subsec. 3.6.1, manifest themselves in the
leading terms of OPE. Thus explicit dependence of vector and scalar
forces on both  $\rho$ and $\rho_s$ may become important.

\subsection{A self-consistent scenario [32,33]}

A rigorous calculation of the contribution of the pion cloud to the
scalar condensate $\kappa(\rho)$ requires the consistent treatment of
the pion dynamics in nuclear matter. The nonlinear term $S(\rho)$ can
be presented as
\begin{equation}
S(\rho)\ =\ \frac{2m^2_\pi}{m_u+m_d}\
\frac{\partial\Sigma_\pi(\rho)}{\partial m^2_\pi}
\end{equation}
with $\Sigma_\pi$ being the pion contribution to the nucleon
self-energy. This can be obtained by using the presentation $S=
dV/dm_q$ (with $V$ standing for the interaction energy) obtained by
Cohen {\em et~al.} \cite{38} and taking into account the pion
contribution only. The self-energy $\Sigma_\pi$ contains the in-medium
pion propagator $D$ which satisfies the Dyson equation
\begin{equation}
D\ =\ D_0+D_0\Pi D\ .
\end{equation}
The propagator $D$ differs from the free propagator $D_0$ due to the
particle-hole excitations described by the polarization operator $\Pi$.
The latter can be expressed through the amplitude of the forward $\pi
N$ scattering (this provides the value $\Pi_0$). The short-range
correlations can be described by means of the Finite Fermi System
Theory (FFST) introduced by Migdal \cite{30}. If only the nucleon-hole
excitations are included, the operator $\Pi_0$ turns to
$\Pi=\Pi_0/(1+g_N\Pi_0)$ with $g_N$ being a FFST constant determined
from the experimental data. The account of long-range correlations
\cite{39} modifies the effective value of $g_N$.

The function $S(\rho)$ thus depends on the effective mass of the
nucleon $m^*(\rho)$ and on the $\pi N$ coupling constant $g^*_{\pi
NN}(\rho)$. The latter can be presented through the fundamental
parameters $f_\pi$ and axial coupling constant $g_A$ by
Goldberger--Treiman relation \cite{40}
\begin{equation}
\frac{g_{\pi NN}}{2m}\ =\ \frac{g_A}{2f_\pi}\ ,
\end{equation}
which can be expanded to the case of the finite density.

On the other hand, the SR provide the dependence of the nucleon
effective mass on the condensate $\kappa(\rho)$. If we succeed in
describing the dependence of $g_A(\rho,\kappa(\rho))$ (the first steps
were made in \cite{20}) and of $f_\pi(\kappa(\rho),\rho)$, we shall
come to the set of self-consistent equations
\begin{equation}
m^*=m^*(\kappa(\rho));\quad f^*_\pi=f^*_\pi(\kappa(\rho)); \quad
g_A=g^*_A(\kappa(\rho)); \quad \kappa=\kappa\left(m^*(\rho),
\frac{g^*_A(\rho)}{f^*_\pi(\rho)}\right).
\end{equation}
The first three equations should originate from the sum rules. The last
one is just the combination of Eqs. (33), (36) and (38). One should add
similar equations for the delta-isobar.

This would enable to describe the baryon parameters in the hadron phase
of the baryon matter up to the point of the chiral phase transition
where $\kappa(\rho)=0$.

\subsection{A sub-plot: Goldstone pions never condense [32,33,41,42]}

The possibility of the "pion condensation" was first discussed by
Migdal \cite{7}. The observation is that at certain value of density
$\rho=\rho_\pi$ the pion propagator presented by Eq. (36) obtains the
pole at the energy $\varepsilon_\pi=0$. This would signal the
degeneracy of the ground state of the system. The ground state contains
an admixture of the oscillations with the quantum numbers of the pions.
The value of $\rho_\pi$ appeared to be sensitive to the values of FFST
constants, being $\rho_\pi\ge2\rho_0$ in most of the hadronic models.

The analysis shows that the singularity of the pion propagator in the
point of the "pion condensation" leads to the divergence of the
function $S(\rho)$ expressed by Eq. (35). This provides
$\kappa(\rho)\to+\infty$ at $\rho\to\rho_\pi$, while the density
increases. On the other hand, $\kappa(0)<0$. Thus $\kappa=0$ at certain
point $\rho_{ch}$ between the zero value and $\rho_\pi$, i.e.
$0<\rho_{ch}<\rho_\pi$. However, since $\kappa(\rho_{ch})=0$, the
chiral symmetry is restored at $\rho=\rho_{ch}$. One cannot expand the
hadronic physics of the densities close to the saturation value
$\rho_0$ to the region $\rho\ge\rho_{ch}$. Thus the "pion condensation"
point cannot be reached in framework of the existing hadronic models.
Anyway, once the chiral symmetry is restored, the pions do not exist as
the Goldstone bosons any more. Thus, there is no "pion condensation" of
the Goldstone pions.

The subject is analyzed in details in the papers \cite{41,42}.
Although not being connected with the SR directly, this result is the
outcome of investigation of the scalar condensate stimulated by studies
of the sum rules.

\section{Summary}

We saw that the first steps in the application of QCD sum rules method
to investigation of the in-medium nucleon parameters are successful.
The effective mass and the single-particle potential energy where
expressed through QCD condensates in the lowest OPE orders. The quark
and gluon condensates are expressed through the observables. The
pion-nucleon $\sigma$-term appeared to determine the value of the
scalar forces. The approach reproduced one of the key points of QHD
picture: the nucleon moves in superposition of scalar and vector fields
which cancel to large extent. The numerical values are consistent with
those of QHD. However the approach does not use the fitting parameters
and avoids the controversial conception of the heavy meson exchange by
the point nucleons.

The approach provides also some new knowledge about the nuclear forces.
These are the anomalous structures of the meson-nucleon vertices. Such
contributions usually are not included in QHD. Another point is the
importance of the scalar channel in the charge-symmetry breaking
forces.

Even the simplified model for the in-medium scalar condensate provides
a possible mechanism of saturation. It is caused by the nonlinear
contribution of the pion cloud to this expectation value. In the
rigorous treatment of the pion dynamics the sum rule for the effective
mass of the nucleon $m^*$ and the expression for the pion contribution
to the condensate form the set of self-consistent equations. The
complete set of the equations should include the sum rules for the
axial coupling constant $g^*_A$ in medium and for the in-medium pion
decay constant $f^*_\pi$. Investigation of the complete set of the
equations is the subject of the future work.

The development of the method requires calculation of the troublesome
four-quark condensates. The calculation of complete set of these
expectation values in framework of the convincing models is in
progress.

I thank M. G. Ryskin, V. A. Sadovnikova and E. M. Levin for fruitful
cooperation during many years. I am indebted to V.M.~Braun, G.E.~Brown,
E.M.~Henley, B.L.~Ioffe, L.~Kisslinger, M.~Rho, E.E.~Saperstein and
C.M.~Shakin for numerous discussions. The work was supported in part by
Deutsche Forschungsgemeinschaft (DFG) --- grant 438/RUS113/595/0-1 and
by Russian Foundation for Basic Research (RFBR) --- grants 0015096610
and 0002-16853.

\end{document}